\input harvmac
\noblackbox

\mathchardef\varGamma="0100
\mathchardef\varDelta="0101
\mathchardef\varTheta="0102
\mathchardef\varLambda="0103
\mathchardef\varXi="0104
\mathchardef\varPi="0105
\mathchardef\varSigma="0106
\mathchardef\varUpsilon="0107
\mathchardef\varPhi="0108
\mathchardef\varPsi="0109
\mathchardef\varOmega="010A

\font\mbm = msbm10
\font\Scr=rsfs10 
\def\bb#1{\hbox{\mbm #1}}

\def\scr#1{\hbox{\Scr #1}}

\def\Mt{{\kern1em\hbox{$\tilde{\kern-1em{\scr M}}$}}}
\def\At{{\kern1em\hbox{$\tilde{\kern-1em{\scr A}}$}}}
\def\Kt{{\kern1em\hbox{$\tilde{\kern-1em{\scr K}}$}}}

\font\sScr=rsfs7 
\def\sscr#1{\hbox{\sScr #1}}

\lref\cargese{
A.~Sagnotti,
``Open Strings And Their Symmetry Groups,''
Cargese Summer Institute on Non-Perturbative Methods in Field Theory,
Cargese, France, 1987
[arXiv:hep-th/0208020];
G.~Pradisi and A.~Sagnotti,
``Open String Orbifolds,''
Phys.\ Lett.\ B216 (1989) 59;
M.~Bianchi and A.~Sagnotti,
``On The Systematics Of Open String Theories,''
Phys.\ Lett.\ B247 (1990) 517;
``Twist Symmetry And Open String Wilson Lines,''
Nucl.\ Phys.\ B361 (1991) 519;
P.~Horava,
``Strings On World Sheet Orbifolds,''
Nucl.\ Phys.\ B327 (1989) 461.
}

\lref\OhtaNQ{
N.~Ohta,
``Cancellation Of Dilaton Tadpoles And Two Loop Finiteness In SO(32) Type I
Superstring,''
Phys.\ Rev.\ Lett.\  59 (1987) 176.
}

\lref\review{
E.~Dudas,
``Theory and phenomenology of type I strings and M-theory,''
Class.\ Quant.\ Grav.\ 17 (2000) R41
[arXiv:hep-ph/0006190];
C.~Angelantonj and A.~Sagnotti,
``Open strings,''
Phys.\ Rept.\ 371 (2002) 1
[Erratum-ibid.\ 376 (2003) 339]
[arXiv:hep-th/0204089].
}

\lref\Polone{
J.~Dai, R.~G.~Leigh and J.~Polchinski,
``New Connections Between String Theories,''
Mod.\ Phys.\ Lett.\ A4 (1989) 2073;
R.~G.~Leigh,
``Dirac-Born-Infeld Action From Dirichlet Sigma Model,''
Mod.\ Phys.\ Lett.\ A4 (1989) 2767.
}

\lref\polchinski{
J.~Polchinski,
``Dirichlet-Branes and Ramond-Ramond Charges,''
Phys.\ Rev.\ Lett.\  75 (1995) 4724
[arXiv:hep-th/9510017].
}

\lref\itoyama{
H.~Itoyama and P.~Moxhay,
``Multiparticle Superstring Tree Amplitudes'',
Nucl.\ Phys.\ B293 (1987) 685.
}

\lref\cai{
J.~Polchinski and Y.~Cai,
``Consistency Of Open Superstring Theories,''
Nucl.\ Phys.\ B296 (1988) 91.
}

\lref\BM{
M.~Bianchi and J.~F.~Morales,
``Anomalies and tadpoles,''
JHEP 0003 (2000) 030
[arXiv:hep-th/0002149].
}

\lref\FS{
W.~Fischler and L.~Susskind,
``Dilaton Tadpoles, String Condensates And Scale Invariance,''
Phys.\ Lett.\ B171 (1986) 383;
``Dilaton Tadpoles, String Condensates And Scale Invariance. 2,''
Phys.\ Lett.\ B173 (1986) 262.
}

\lref\DNPS{
E.~Dudas, G.~Pradisi, M.~Nicolosi and A.~Sagnotti,
``On tadpoles and vacuum redefinitions in string theory,''
arXiv:hep-th/0410101.
}

\lref\typezero{
L.~J.~Dixon and J.~A.~Harvey,
``String Theories In Ten-Dimensions Without Space-Time Supersymmetry,''
Nucl.\ Phys.\ B274 (1986) 93;
N.~Seiberg and E.~Witten,
``Spin Structures In String Theory,''
Nucl.\ Phys.\ B276 (1986) 272.
}

\lref\yassen{
D.~Fioravanti, G.~Pradisi and A.~Sagnotti,
``Sewing constraints and nonorientable open strings,''
Phys.\ Lett.\ B321 (1994) 349
[arXiv:hep-th/9311183];
G.~Pradisi, A.~Sagnotti and Y.~S.~Stanev,
``Planar duality in SU(2) WZW models,''
Phys.\ Lett.\ B354 (1995) 279
[arXiv:hep-th/9503207];
``The Open descendants of nondiagonal SU(2) WZW models,''
Phys.\ Lett.\ B356 (1995) 230
[arXiv:hep-th/9506014];
``Completeness Conditions for Boundary Operators in 2D Conformal Field Theory,''
Phys.\ Lett.\ B381 (1996) 97
[arXiv:hep-th/9603097].
}

\lref\bert{
L.~R.~Huiszoon, A.~N.~Schellekens and N.~Sousa,
``Klein bottles and simple currents,''
Phys.\ Lett.\ B470 (1999) 95
[arXiv:hep-th/9909114].
L.~R.~Huiszoon and A.~N.~Schellekens,
``Crosscaps, boundaries and T-duality,''
Nucl.\ Phys.\ B584 (2000) 705
[arXiv:hep-th/0004100].
J.~Fuchs, L.~R.~Huiszoon, A.~N.~Schellekens, C.~Schweigert and J.~Walcher,
``Boundaries, crosscaps and simple currents,''
Phys.\ Lett.\ B495 (2000) 427
[arXiv:hep-th/0007174].
}

\lref\surprises{
A.~Sagnotti,
``Some properties of open string theories,''
arXiv:hep-th/9509080;
``Surprises in open-string perturbation theory,''
Nucl.\ Phys.\ Proc.\ Suppl.\  56B (1997) 332
[arXiv:hep-th/9702093].
}

\lref\alvarez{
L.~Alvarez-Gaume, P.~H.~Ginsparg, G.~W.~Moore and C.~Vafa,
``An $O(16) \times O(16)$ Heterotic String,''
Phys.\ Lett.\ B171 (1986) 155.
}

\lref\susyzeroB{
C.~Angelantonj,
``Non-tachyonic open descendants of the 0B string theory,''
Phys.\ Lett.\ B444 (1998) 309
[arXiv:hep-th/9810214];
R.~Blumenhagen, A.~Font and D.~Lust,
``Non-supersymmetric gauge theories from D-branes in type 0 string  theory,''
Nucl.\ Phys.\ B560 (1999) 66
[arXiv:hep-th/9906101];
K.~F\"orger,
``On non-tachyonic $Z_N \times Z_M$ orientifolds of type 0B string theory,''
Phys.\ Lett.\ B469 (1999) 113
[arXiv:hep-th/9909010].
}

\lref\kumar{
R.~Blumenhagen and A.~Kumar,
``A note on orientifolds and dualities of type 0B string theory,''
Phys.\ Lett.\ B464 (1999) 46
[arXiv:hep-th/9906234].
}

\lref\malda{
J.~M.~Maldacena,
``The large N limit of superconformal field theories and supergravity,''
Adv.\ Theor.\ Math.\ Phys.\  2 (1998) 231
[Int.\ J.\ Theor.\ Phys.\ 38 (1999) 1113]
[arXiv:hep-th/9711200];
S.~S.~Gubser, I.~R.~Klebanov and A.~M.~Polyakov,
``Gauge theory correlators from non-critical string theory,''
Phys.\ Lett.\ B428 (1998) 105
[arXiv:hep-th/9802109];
E.~Witten,
``Anti-de Sitter space and holography,''
Adv.\ Theor.\ Math.\ Phys.\  2 (1998) 253
[arXiv:hep-th/9802150].
See also 
O.~Aharony, S.~S.~Gubser, J.~M.~Maldacena, H.~Ooguri and Y.~Oz,
``Large N field theories, string theory and gravity,''
Phys.\ Rept.\  323 (2000) 183
[arXiv:hep-th/9905111]
 for a review.
 }

\lref\adi{
C.~Angelantonj and A.~Armoni,
``Non-tachyonic type 0B orientifolds, non-supersymmetric gauge theories  and
cosmological RG flow,''
Nucl.\ Phys.\ B578 (2000) 239
[arXiv:hep-th/9912257].
C.~Angelantonj and A.~Armoni,
``RG flow, Wilson loops and the dilaton tadpole,''
Phys.\ Lett.\ B482 (2000) 329
[arXiv:hep-th/0003050].
}

\lref\adiven{
For a review, see A.~Armoni, M.~Shifman and G.~Veneziano,
``From super-Yang-Mills theory to QCD: Planar equivalence and its implications,''
arXiv:hep-th/0403071.
}

\lref\KachruHD{
S.~Kachru, J.~Kumar and E.~Silverstein,
``Vacuum energy cancellation in a non-supersymmetric string,''
Phys.\ Rev.\ D59 (1999) 106004[arXiv:hep-th/9807076].
}

\lref\HarveyRC{
J.~A.~Harvey,
``String duality and non-supersymmetric strings,''
Phys.\ Rev.\ D59 (1999) 026002
[arXiv:hep-th/9807213].
}

\lref\KachruPG{
S.~Kachru and E.~Silverstein,
``On vanishing two loop cosmological constants in non-supersymmetric
strings,'' JHEP 9901 (1999) 004
[arXiv:hep-th/9810129];
R.~Iengo and C.~J.~Zhu,
``Evidence for non-vanishing cosmological constant in non-SUSY superstring
models,''
JHEP 0004 (2000) 028
[arXiv:hep-th/9912074].
}

\lref\BlumenhagenUF{
R.~Blumenhagen and L.~G\"orlich,
``Orientifolds of non-supersymmetric, asymmetric orbifolds,''
Nucl.\ Phys.\ B551 (1999) 601
[arXiv:hep-th/9812158].
}

\lref\AngelantonjGM{
C.~Angelantonj, I.~Antoniadis and K.~Forger,
``Non-supersymmetric type I strings with zero vacuum energy,''
Nucl.\ Phys.\ B555 (1999) 116
[arXiv:hep-th/9904092].
}

\lref\BL{
M.~Berkooz, M.~R.~Douglas and R.~G.~Leigh,
``Branes intersecting at angles,''
Nucl.\ Phys.\ B480 (1996) 265
[arXiv:hep-th/9606139].
}

\lref\magnetic{
C.~Bachas,
``A Way to break supersymmetry,''
arXiv:hep-th/9503030.
 R.~Blumenhagen, L.~G\"orlich, B.~K\"ors and D.~L\"ust,
``Noncommutative compactifications of type I strings on tori with  magnetic background flux,''
JHEP 0010 (2000) 006
[arXiv:hep-th/0007024];
C.~Angelantonj, I.~Antoniadis, E.~Dudas and A.~Sagnotti,
``Type-I strings on magnetised orbifolds and brane transmutation,''
Phys.\ Lett.\ B489 (2000) 223
[arXiv:hep-th/0007090].
}

\lref\dieter{
A.~M.~Uranga,
``Chiral four-dimensional string compactifications with intersecting D-branes,''
Class.\ Quant.\ Grav.\  20 (2003) S373
[arXiv:hep-th/0301032];
D.~Cremades, L.~E.~Ibanez and F.~Marchesano,
``More about the standard model at intersecting branes,''
arXiv:hep-ph/0212048;
D.~L\"ust,
``Intersecting brane worlds: A path to the standard model?,''
Class.\ Quant.\ Grav.\  21 (2004) S1399
[arXiv:hep-th/0401156].
}

\lref\stability{
C.~Angelantonj, R.~Blumenhagen and M.~R.~Gaberdiel,
``Asymmetric orientifolds, brane supersymmetry breaking and non-BPS  branes,''
Nucl.\ Phys.\ B589 (2000) 545
[arXiv:hep-th/0006033].
}

\lref\melvin{
C.~Angelantonj, E.~Dudas and J.~Mourad,
``Orientifolds of string theory Melvin backgrounds,''
Nucl.\ Phys.\ B637 (2002) 59
[arXiv:hep-th/0205096];
C.~Angelantonj,
``Rotating D-branes and O-planes,''
Fortsch.\ Phys.\  51 (2003) 646
[arXiv:hep-th/0212066].
}

\lref\moore{
D.~Kutasov, J.~Marklof and G.~W.~Moore,
``Melvin models and diophantine approximation,''
arXiv:hep-th/0407150.
}

\lref\SSi{
J.~Scherk and J.~H.~Schwarz,
``How To Get Masses From Extra Dimensions,''
Nucl.\ Phys.\ B153 (1979) 6;
``Spontaneous Breaking Of Supersymmetry Through Dimensional Reduction,''
Phys.\ Lett.\ B82 (1979) 60.
}

\lref\SSiii{
R.~Rohm,
``Spontaneous Supersymmetry Breaking In Supersymmetric String Theories,''
Nucl.\ Phys.\ B237 (1984) 553;
C.~Kounnas and M.~Porrati,
``Spontaneous Supersymmetry Breaking In String Theory,''
Nucl.\ Phys.\ B310 (1988) 355;
S.~Ferrara, C.~Kounnas, M.~Porrati and F.~Zwirner,
``Superstrings With Spontaneously Broken Supersymmetry And Their
Effective Theories,''
Nucl.\ Phys.\ B318 (1989) 75.
}

\lref\taylor{
H.~Itoyama and T.~R.~Taylor,
``Supersymmetry Restoration In The Compactified $O(16) \times O(16)'$
Heterotic String Theory,'' Phys.\ Lett.\ B186 (1987) 129.
}

\lref\ADS{
I.~Antoniadis, E.~Dudas and A.~Sagnotti,
``Supersymmetry breaking, open strings and M-theory,''
Nucl.\ Phys.\ B544 (1999) 469
[arXiv:hep-th/9807011];
I.~Antoniadis, G.~D'Appollonio, E.~Dudas and A.~Sagnotti,
``Partial breaking of supersymmetry, open strings and M-theory,''
Nucl.\ Phys.\ B553 (1999) 133
[arXiv:hep-th/9812118];
A.~L.~Cotrone,
``A $Z_2 \times Z_2$ orientifold with spontaneously broken supersymmetry,''
Mod.\ Phys.\ Lett.\ A 14 (1999) 2487
[arXiv:hep-th/9909116];
P.~Anastasopoulos, A.~B.~Hammou and N.~Irges,
``A class of non-supersymmetric open string vacua,''
Phys.\ Lett.\ B581 (2004) 248
[arXiv:hep-th/0310277].
}

\lref\BSB{
S.~Sugimoto,
 ``Anomaly cancellations in type I ${\rm D}9$-$\overline{{\rm D}9}$ system and the USp(32)  string theory'',
Prog.\ Theor.\ Phys.\  102 (1999) 685
[arXiv:hep-th/9905159];
I.~Antoniadis, E.~Dudas and A.~Sagnotti,
``Brane supersymmetry breaking,''
Phys.\ Lett.\ B464 (1999) 38
[arXiv:hep-th/9908023].
}

\lref\BSBtwo{
C.~Angelantonj, I.~Antoniadis, G.~D'Appollonio, E.~Dudas and A.~Sagnotti,
``Type I vacua with brane supersymmetry breaking,''
Nucl.\ Phys.\ B572 (2000) 36
[arXiv:hep-th/9911081].
}

\lref\bianchi{
M. Bianchi, PhD Thesis (1992);
A.~Sagnotti,
``Anomaly cancellations and open string theories,''
arXiv:hep-th/9302099;
G.~Zwart,
``Four-dimensional $N = 1$ $Z_N \times Z_M$ orientifolds,''
Nucl.\ Phys.\ B526 (1998) 378
[arXiv:hep-th/9708040].
}

\lref\ej{
E.~Dudas and J.~Mourad,
``Consistent gravitino couplings in non-supersymmetric strings,''
Phys.\ Lett.\ B514 (2001) 173
[arXiv:hep-th/0012071];
G.~Pradisi and F.~Riccioni,
``Geometric couplings and brane supersymmetry breaking,''
Nucl.\ Phys.\ B615 (2001) 33
[arXiv:hep-th/0107090].
For a review of the main features of BSB and its field theory implications, see
C.~Angelantonj,
``Aspects of supersymmetry breaking in open-string models,''
Fortsch.\ Phys.\ 50 (2002) 735.
}

\lref\ldim{
I.~Antoniadis, ``A Possible New Dimension at a Few TeV,''
Phys. Lett. B246 (1990)~377;
N.~Arkani-Hamed, S.~Dimopoulos and G.~Dvali,
``The Hierarchy Problem and New Dimensions at a Millimetre,''
Phys. Lett. B429 (1998) 263, [arXiv:hep-ph/9803315];
I.~Antoniadis, N.~Arkani-Hamed, S.~Dimopoulos and G.~R.~Dvali,
``New dimensions at a millimetre to a Fermi and superstrings at a TeV,''
Phys.\ Lett.\ B436 (1998) 257
[arXiv:hep-ph/9804398].
}

\lref\tristan{
M.~Borunda, M.~Serone and M.~Trapletti,
``On the quantum stability of type IIB orbifolds and orientifolds 
with Scherk-Schwarz SUSY breaking,''
Nucl.\ Phys.\ B 653 (2003) 85
[arXiv:hep-th/0210075];
I.~Antoniadis, K.~Benakli, A.~Laugier and T.~Maillard,
``Brane to bulk supersymmetry breaking and radion force at micron
distances,'' Nucl.\ Phys.\ B662 (2003) 40
[arXiv:hep-ph/0211409].
}

\lref\discrete{
C.~Angelantonj and R.~Blumenhagen,
``Discrete deformations in type I vacua,''
Phys.\ Lett.\ B473 (2000) 86
[arXiv:hep-th/9911190].
}

\lref\toroidal{
M.~Bianchi, G.~Pradisi and A.~Sagnotti,
``Toroidal compactification and symmetry breaking in open string theories,''
Nucl.\ Phys.\ B376 (1992) 365;
E.~Witten,
``Toroidal compactification without vector structure,''
JHEP 9802 (1998) 006
[arXiv:hep-th/9712028];
M.~Bianchi,
``A note on toroidal compactifications of the type I superstring and  other
superstring vacuum configurations with 16 supercharges,''
Nucl.\ Phys.\ B528 (1998) 73
[arXiv:hep-th/9711201].
}

\lref\Borbifold{
C.~Angelantonj,
``Comments on open-string orbifolds with a non-vanishing $B_{ab}$,''
Nucl.\ Phys.\ B566 (2000) 126
[arXiv:hep-th/9908064].
}

\lref\systematics{ 
M. Bianchi and A. Sagnotti in \cargese.
}

\lref\generalised{
A.~Sagnotti,
``A Note on the Green-Schwarz mechanism in open string theories,''
Phys.\ Lett.\ B294 (1992) 196
[arXiv:hep-th/9210127].
}

\lref\Bmagnetic{
C.~Angelantonj and A.~Sagnotti,
``Type-I vacua and brane transmutation,''
arXiv:hep-th/0010279;
R.~Blumenhagen, B.~Kors and D.~Lust,
``Type I strings with F- and B-flux,''
JHEP 0102 (2001) 030
[arXiv:hep-th/0012156].
}

\lref\suppressing{
C.~Angelantonj and I.~Antoniadis,
``Suppressing the cosmological constant in non-supersymmetric type I strings,''
Nucl.\ Phys.\ B676 (2004) 129
[arXiv:hep-th/0307254].
}

\lref\higher{
M.~Bianchi and A.~Sagnotti,
``Open Strings And The Relative Modular Group,''
Phys.\ Lett.\ B231 (1989) 389;
I.~Antoniadis and T.~R.~Taylor,
``Topological masses from broken supersymmetry,''
Nucl.\ Phys.\ B695 (2004) 103
[arXiv:hep-th/0403293].
}

\lref\cardella{
C.~Angelantonj and M.~Cardella,
``Vanishing perturbative vacuum energy in non-supersymmetric orientifolds,''
Phys.\ Lett.\ B595 (2004) 505
[arXiv:hep-th/0403107].
}


\Title{\vbox{\rightline{\tt hep-th/0411085}
\rightline{LMU-TPS 04/11}}}
{\vbox{\centerline{Open Strings and Supersymmetry Breaking}}}

\centerline{Carlo Angelantonj}
\medskip
\centerline{\it Department f\"ur Physik, Ludwig-Maximilians-Universi\"at, M\"unchen}
\centerline{\it Theresienstr. 37, D80333 M\"unchen}

\vskip 0.8 in

\centerline{{\bf Abstract}}

We review several mechanisms for supersymmetry breaking in orientifold models. In particular, we focus on non-supersymmetric open-string realisations that correspond to consistent flat-space solutions of the classical equations of motion. In these models, the one-loop vacuum energy can typically fixed by the size of the compact extra dimensions, and can thus be tuned to extremely small values if enough extra dimensions are large.

\Date{November, 2004}

\newsec{A glimpse at orientifold constructions}

Type I models have become the subject of an intense activity during the last few years, since their perturbative definition offers interesting new possibilities for low-energy phenomenology. Their consistency and a number of their most amusing features may be traced back to the relation to suitable ``parent'' models of oriented closed strings \refs{\cargese,\review}, from which their spectra can be derived. In this procedure, a special role is played by ``tadpole conditions'' for Ramond-Ramond (R-R) and Neveu-Schwarz-Neveu-Schwarz (NS-NS) massless states. 
One of the most amusing features of orientifold constructions is the different origin of gravitational and gauge interactions. Although this observation might seem in contrast with the old dream of unifying geometry and matter, it actually allows for more generic vacuum configurations with interesting implications. On the one hand, gravity originates from closed strings, and as such permeates the whole ten-dimensional space time. On the other hand, gauge interactions are associated to open strings whose free ends live on D$p$-branes \Polone, $p+1$ dimensional hyper manifolds embedded in the ten-dimensional bulk. In particular, if we denote by $x^\mu$ ($y^a$) the coordinates along (transverse to) the world-volume of the D-branes, this simple observation implies for example that the full metric tensor and the space-time gauge field are of the form 
\eqn\dbranes{
g_{MN} = g_{MN} (x^\rho , y^a) \,, \qquad A_\mu = A_\mu (x^\rho) \,,
}
and, as we shall see, this fact has dramatic consequences in orientifold constructions. 

Given this geometric description of orientifold constructions, one can associate a more physical interpretation to tadpole conditions. While R-R tadpoles are to be regarded as {\it global} neutrality conditions for R-R charges \polchinski, NS-NS tadpoles ensure that the configurations of D-branes and O-planes (a sort of rigid mirrors that revert the orientation of closed and open strings) be {\it globally} massless. 
In supersymmetric models these two conditions are related by supersymmetry transformations, and thus the vanishing of R-R tadpoles naturally implies that the NS-NS ones vanish as well\footnote{${}^\dagger$}{See \itoyama\ and \OhtaNQ\ for a discussion on the vanishing of NS-NS tadpoles}. However, it is worth stressing here that they have completely different consequences: while the R-R tadpole conditions are required by gauge invariance and 
their violation is linked to the emergence of irreducible gauge and gravitational anomalies \refs{\cai,\BM}, NS-NS tadpoles are not associated to any inconsistency, and thus in principle may be relaxed. In doing so, however, one perturbs the background geometry 
\FS, with the end result that full fledged string theory calculations are more difficult to perform \DNPS. 

\newsec{Non tachyonic type 0 vacua in various dimensions}

Looking for non supersymmetric vacua, type 0 strings \typezero\ offer a natural arena. Typically, these theories include in their spectrum tachyonic modes (both in the closed and in the open sector) that induce instabilities in the vacuum. However, it was shown in \refs{\yassen,\bert} that orientifold constructions allow for different choices of projections and in particular in \surprises\ a tachyon-free descendant of the type 0B theory was built. Together with non-tachyonic non-supersymmetric heterotic strings \alvarez, this is a notable example of a classically stable string vacuum without supersymmetry.

A natural question is then whether these properties will survive upon compactification on non trivial manifolds. While surprises are not expected in simple toroidal reductions, new interesting features emerge when orbifolds are considered. Since the ten-dimensional parent strings are not supersymmetric to begin with, one is now entitled to use both supersymmetry-preserving \susyzeroB\ and supersymmetry-breaking twists \kumar. In the case of $T^4 /\bb{Z}_N$ and $T^6 /\bb{Z}_N$ compactifications, an exhaustive analysis has revealed that, for generic $N$, aside from the surviving untwisted tachyon of the parent closed oriented theory, new complex twisted tachyons are typically present in the spectrum of light excitations \refs{\susyzeroB,\kumar}.  As a result, a generic orientifold projection can only make the twisted tachyons real, and thus classical instabilities are always present. The only exception is given by the supersymmetry-preserving $T^4 /\bb{Z}_2$, $T^6 /\bb{Z}_3$ and $T^6 /\bb{Z}_2 \times \bb{Z}_2$ and supersymmetry-breaking $T^6/\bb{Z}_2$ cases, where no tachyons emerge in the twisted sectors. One can then properly deform the orientifold projection as in \surprises\ and build new lower-dimensional vacua free from untwisted and twisted tachyonic modes \susyzeroB.
One should stress here that, although these vacua are tachyon-free, the impossibility of cancelling their NS-NS tadpoles and the generation of a one-loop cosmological constant inevitably destabilise the classical vacuum. 

Type 0 theories and their D-branes have also triggered some activity in the context of 
gauge/gravity dualities \malda\ for non-supersymmetric non-conformal theories.
While in oriented type 0B strings the presence of the tachyon complicates the gravity description and leads to instabilities in the dual strongly coupled field theory, a sensible holographic description of non-supersymmetric gauge theories was proposed in \adi. Resorting to the non-tachyonic orientifolds of \surprises, the gauge theory in the open-string sector turns out to be conformal in the planar large-$N$ limit, and therefore several results can be argued from the parent ${\cal N}=4$ theory. In particular, the leading (planar) geometry remains $AdS_5 \times S^5$ with a constant dilaton field. For finite $N$, however, the gauge theory is no longer conformal and new features are expected in the gravity description. Indeed, the lack of conformal invariance translates on the string side into the presence of a non-vanishing dilaton tadpole at the disk level that induces deviations from the $AdS_5 \times S^5$ geometry. In addition, the dilaton starts running consistently with the RG-determined behaviour for the gauge coupling on the field theory side. Moreover, the back-reaction in the classical geometry predicts a quark-anti-quark potential interpolating between a (logarithmically running) Coulomb phase and a confining phase in the IR, as expected from the field theory analysis \adi. More recently this study has been extended to obtain exact results in non-supersymmetric gauge theories \adiven.

\newsec{Breaking supersymmetry spontaneously}

Compared to explicit breakings of supersymmetry by suitable choices of modular invariant partition functions and/or compactification manifolds, the spontaneous breaking has more 
appealing properties since at times it allows one to attain a better control on  radiative corrections of masses and couplings.

Independently of the specific mechanism at work, whenever supersymmetry is spontaneously broken at a scale $M_{\rm sb}$ (depending on the specific model one is considering), the (tree-level) mass splitting in a given super multiplet and the (one-loop induced) cosmological constant are typically both determined by $M_{\rm sb}$:
\eqn\cosmgau{
\varLambda \sim M_{\rm sb}^4 \,, \qquad \delta M \sim M_{\rm sb}\,.
}
One of the outstanding problems in string theory, as in any quantum theory including gravity, is to understand how a small cosmological constant can be accompanied by the generation of appropriate gaugino masses. This problem became much more severe after recent observations suggested a non-vanishing vacuum energy corresponding to a new energy scale, far smaller than every other scale in the physics of fundamental interactions, $\varLambda_{\rm obs} = E_\varLambda^4$, with $E_\varLambda \sim 10^{-4} \, {\rm eV}$. Similarly, experiments in Particle Physics suggest that gauginos should be heavier than a few TeV. Given the expressions in eq. \cosmgau, two distinct mechanisms for breaking supersymmetry should be combined to successfully disentangle the cosmological constant scale from the gaugino mass scale.

Type II string models with a cosmological constant possibly vanishing in
perturbation theory were studied in refs. \refs{\KachruHD,\HarveyRC}.
Their main feature is a Fermi-Bose degenerate spectrum,
leading to an automatic vanishing of the one-loop vacuum energy.
Aside from the question of higher-loop corrections \KachruPG,
their main defect is that the non-abelian gauge
sector, appearing at particular singular points of the compactification
manifold, or on appropriate D-brane collections, is always
supersymmetric \refs{\BlumenhagenUF, \AngelantonjGM}.
Thus, it is questionable whether such constructions can
accommodate gauge degrees of freedom with large supersymmetry-breaking
mass splittings without spoiling the vanishing of the vacuum energy.

Among the mechanisms to break supersymmetry in the gauge sector, intersecting brane models \BL, T-dual to magnetic field backgrounds \magnetic, have proved to be a natural setting to realise Standard-Model-like patterns of gauge symmetries and matter fields within string theory (see \dieter\ for a review and references therein). However, they are generally plagued by tachyonic instabilities, and only in few cases the fate of these unstable configurations has been studied in some detail \stability.
Moreover, in standard realisations it seems quite difficult to give masses to the superpartners of the $SU(3) \times SU(2) \times U(1)$ gauge bosons, and typically a sizeable vacuum energy determined by the intersection angles (or by the strength of the background magnetic fields) is generated already at the disk level \magnetic. 
These constructions can be extended to the case of magnetic backgrounds for 
the gravi-photon, corresponding to strings fluctuating in Melvin spaces. In this case the breaking of supersymmetry is felt also by the closed-string sector, and a T-dual description would include orientifold planes at generic angles \melvin. Although this scenario has not found a direct application in string phenomenology, it provides an interesting example of conformal field theories with non-compact target spaces associated to orbifolds with irrational twists, and also exhibits some amusing arithmetic properties \moore.

The Scherk-Schwarz mechanism provides an elegant realisation of supersymmetry breaking by compactification in field theory \SSi. In the simplest case of circle compactification, it amounts to allowing the higher dimensional fields to be periodic around the circle up to an R-symmetry transformation. The Kaluza-Klein momenta of the various fields are correspondingly shifted proportionally to their R charges, and modular invariance dictates the extension of this mechanism to the full perturbative spectrum in models of oriented closed strings \SSiii. As a result the gravitini get masses inversely proportional to the compactification radius
$$
m_{3/2} \sim R^{-1} \,,
$$
while the breaking of supersymmetry is accompanied by a one-loop vacuum energy that reproduces the behaviour familiar from field theory
$$
\varLambda_{\rm SS} \sim R^{-4} \,,
$$
aside from additional terms exponentially suppressed with $R$ originating from the twisted sector \taylor. When open strings are present, one has to distinguish between the two cases of Scherk-Schwarz deformations transverse or longitudinal to the world-volume of the branes \ADS. In fact, as a result of \dbranes\ in the former case the open-string fields do not depend on the coordinates of the extra dimension, and therefore are not affected by the deformation. In this scenario, termed in \ADS\ ``M-theory breaking'', the D-brane excitations stay supersymmetric (at least to lowest order) and the gaugino mass is identically vanishing, $m_{1/2} =0$. In the latter case, instead, the R charges determine the masses of the fields and $m_{1/2} \sim R^{-1}$. 

Finally, Brane Supersymmetry Breaking \BSB\ is a purely stringy mechanism that to lowest order affects only the open-string excitations. The supersymmetric bulk is coupled to non-supersymmetric branes, where the mass splitting is set by the string scale itself,
$m_{1/2} \sim M_{\rm s}$. However, Brane Supersymmetry Breaking yields a non-vanishing contribution to the cosmological constant already at the disk level, due to the impossibility of cancelling NS-NS tadpoles, with $\varLambda \sim M_{\rm s}^4$.
In a number of interesting examples, brane supersymmetry breaking provides a solution \BSBtwo\ to an old puzzle in the construction of open-string models, where some tadpole conditions were long known to allow apparently no consistent solution \bianchi. At the level of the low-energy effective action the apparently broken supersymmetry is linearly realised in the open-string sector and the a singlet spinor, always present in these models, plays the role of the goldstino \ej.

\newsec{Scales in orientifold models}

We have long been accustomed to accepting the fate that typical string effects are confined to very high energies. This is directly implied by the often implicit identification of the string scale $M_{\rm s} = \ell_{\rm s}^{-1}$ with the Planck scale $M_{\rm Pl}$, motivated by the experience with the weakly coupled heterotic strings, where gravitational and gauge interactions are associated to excitations of closed strings only. It is amazing that the different nature of gravity and gauge forces in orientifold models, together will the built-in observation that they generally propagate in different space-time directions, as recalled in eq. \dbranes, allows one to decouple $\ell_{\rm s}$ from $\ell_{\rm Pl}$. Indeed, the schematic low energy effective action of a generic orientifold model in the presence of a D$p$ brane reads
$$
S = \int [ d^{10} x ] \, {1\over \ell_{\rm s}^8 \, g_{\rm s}^2} R + \int [d^{p+1} x] \, {1\over \ell_{\rm s}^{p-3} \, g_{\rm s}} F^2 \,,
$$
with $g_{\rm s}$ the string coupling constant, and $R$ and $F^2$ the familiar Ricci scalar and the kinetic term for gauge fields. Upon compactification to four dimensions, the Planck length and the gauge coupling constant 
\eqn\couplings{
{1\over \ell_{\rm Pl}^2} = {V_\| \, V_\perp \over  \ell_{\rm s}^8 \, g_{\rm s}^2} \,,
\qquad {1\over g_{\rm YM}^2} = {V_\| \over \ell_{\rm s}^{p-3} \, g_{\rm s}} \,,
}
are related to the volumes $V_\|$ and $V_\perp$ parallel and orthogonal to the branes as well as to the string length $\ell_{\rm s}$ and to the string coupling constant $g_{\rm s}$. Combining eqs. \couplings, one can arrive at the key expressions
\eqn\reduced{
M_{\rm Pl}^2 = {1\over g_{\rm YM}^4 v_\|} M_{\rm s}^{2+n} R^n_\perp \,, \qquad
g_{\rm s} = g^2_{\rm YM} v_\| \,,
}
with $n=9-p$ the number of dimensions, all of size $R_\perp$, orthogonal to the branes and $v_\| = V_\| / \ell_{\rm s}^{p-3}$ the longitudinal volume measured in string units. It should then be clear that in a weakly coupled type I string $v_\| \sim 1$, while the string scale and the size of the transverse directions are completely undetermined, though correlated through \reduced. For instance for $M_{\rm s} \sim 1\, {\rm TeV}$ the size of the extra dimensions $R_\perp$ can in principle vary from $10^8\, {\rm km}$, to $0.1\, {\rm mm}$ down to $0.1\, {\rm fm}$ for $n=1,2$ or 6 transverse directions \ldim. Aside from the $n=1$ case, clearly excluded, all other cases are actually consistent with observations.

\newsec{Suppressing the cosmological constant}

It is very suggestive, if not a simple numerical coincidence, that for $n=2$ the size of the transverse dimensions is of the same order of magnitude as the observed cosmological constant scale, $R_\perp \sim E_\varLambda^{-1}$. In fact, as we have seen, the behaviour $\varLambda \sim R_\perp^{-4}$ is typical of models where supersymmetry is broken by Scherk-Schwarz deformations. However, in this case mass splittings are at most of order $R_\perp^{-1}$, and thus too small by several orders of magnitude. Similarly, in models featuring Brane Supersymmetry Breaking the gaugino mass, determined by the string scale itself, can be naturally tuned to a few TeV, consistently with data from particle accelerators. 
It would thus be tempting to combine Scherk-Schwarz reductions transverse to the branes with brane supersymmetry breaking in order to disentangle the cosmological constant scale and the gaugino mass scale, and to tune them to experimentally acceptable values. However, a naive combination of these two effects would appear to spoil the value of the vacuum energy, since
\eqn\vacuum{
\varLambda (R) \sim (n^c_{\rm B} - n^c_{\rm F}) {1\over R^4} + c_1 (n^o_{\rm B} - n^o_{\rm F}) M_{\rm s}^4 + c_2 (n^o_{\rm B} - n^o_{\rm F}) {M_{\rm s}^{6-n} \over R^{n-2}} + {\scr O} \left( e^{-M_{\rm s}^2 R^2} \right)
}
would receive contributions of order $M_{\rm s}^4 \sim ({\rm few\ TeV})^4$ from the brane supersymmetry breaking mechanism, unless the open-string spectrum is Fermi-Bose degenerate. 
In one possible realisation, this degeneracy might be thought of as emerging from a supersymmetric theory where the super partners have been displaced appropriately in position space.

In order to achieve such a Fermi-Bose degenerate open-string spectrum one turns on discrete values for the NS-NS $B_{ab}$ along the compact directions. The presence of such a background,
or of similar discrete vacuum expectation values of fields projected out by the world-sheet parity \discrete, reverts the nature of some of the orientifold planes \toroidal, and therefore fewer numbers of D-branes are required to cancel tadpoles. As a result, in the open-string sector the rank of the gauge group is reduced proportionally to the rank of the non-vanishing $B_{ab}$, and moreover symplectic groups can be continuously connected to orthogonal ones \toroidal. In orbifold compactifications these phenomena are accompanied by a modified structure of the fixed points, that in the presence of a quantised $B_{ab}$ arrange themselves into multiplets and induce different projections in the twisted sector \Borbifold. As a result, in six-dimensional ${\scr N} = (1,0)$ compactifications variable numbers of tensor multiplets are present in the closed unoriented spectrum, thus  providing a geometrical description of the rational constructions first presented in \systematics. The presence of these additional tensors is actually crucial for obtaining consistent vacuum configurations, since they play a significant role in a generalised Green-Schwarz mechanism for anomaly cancellation \generalised.
Discrete values for the NS-NS $B$-field are of crucial importance also in compactifications on magnetised backgrounds, where odd numbers of families of chiral matter are then allowed \Bmagnetic. 

The simplest string theory construction with a Fermi-Bose degenerate spectrum  \suppressing\ corresponds to the M-theory breaking model of \ADS\ compactified on an additional $T^2$ that is permeated with a quantised $B_{ab}$. To be more concrete, the one-loop torus, Klein, annulus, and M\"obius-strip amplitudes 
$$
\eqalign{
{\scr T} =& {\textstyle{1\over 2}} \left[ (V_8 \bar V_8 + S_8 \bar S_8 ) \varGamma^{(1,1)}_{m,2n} +
(O_8 \bar O_8 + C_8 \bar C_8) \varGamma^{(1,1)}_{m,2n+1} \right.
\cr
& \left. - (V_8 \bar S_8 + S_8 \bar V_8 ) \varGamma^{(1,1)}_{m+{1\over 2}, 2n}
- (O_8 \bar C_8 + C_8 \bar O_8 ) \varGamma^{(1,1)}_{m+{1\over 2}, 2n+1}
\right] \varGamma^{(2,2)} (B) \,,
}
$$
$$
{\scr K} = {\textstyle{1\over 2}} \left[ (V_8 - S_8) W^{(1)}_{2n}
+ (O_8 - C_8) W^{(1)}_{2n+1} \right] W^{(2)}_{(2n^7 , 2 n^8)} \,,
$$
$$
{\scr A} = {\textstyle{1\over 2}} \left( N_{\rm D}^2 + N_{\bar{\rm D}}^2
\right) (V_8 - S_8) W^{(1)}_{n} W^{(2)}_{(n^7 , n^8)}
+ N_{\rm D} N_{\bar{\rm D}} (O_8 - C_8 ) W^{(1)}_{n+{1\over 2}}
W^{(2)}_{(n^7 + {1\over 2}, n^8 )} \,,
$$
and
$$
\eqalign{
{\scr M} =& - {\textstyle{1\over 2}} \left( V_8 + (-1)^n S_8 \right)
W^{(1)}_n \left[ (N_{\rm D} + N_{\bar{\rm D}} ) W^{(2)}_{(n^7 , 2 n^8 + 1)}
\right. 
\cr
& \left. - (N_{\rm D} - N_{\bar{\rm D}} ) (-1)^{n^7} W^{(2)}_{(n^7 , 2 n^8)}
\right] \,,
\cr}
$$
encode all the information about the geometry and the spectrum of the vacuum configuration (see \suppressing\ for details). 
Supersymmetry is clearly broken {\it \`a la} Scherk-Schwarz in the bulk, while the D-brane massless excitations comprise the Kaluza-Klein reduction of ten-dimensional gauge bosons with gauge group ${\rm USp} (8) \times {\rm SO} (8)$, together with fermions in the $(28,1)+(1,36)$ representations, as a result of Brane Supersymmetry Breaking. The two gauge group factors, and thus the two sets of branes, are distributed at different points along the compact directions where suitable orientifold planes are located. This open-string spectrum is clearly Fermi-Bose degenerate, although it is not supersymmetric.

The one-loop vacuum energy is then
$$
\eqalign{
\varLambda (R) =& \int_{\sscr F} {d^2 \tau \over \tau_2^{9/2}} \, {{\scr T} (R) \over |\eta |^{10}} + \int_0^\infty {d\tau_2 \over \tau_2^{9/2}} {{\scr K} (R) \over \eta^5}
\cr
& + \int_0^\infty {d\tau_2 \over \tau_2^{9/2}} {{\scr A} (R) \over \eta^5}
+ \int_0^\infty {d\tau_2 \over \tau_2^{9/2}} {{\scr M} (R) \over \eta^5} \,,
\cr}
$$
and, aside from the power-low behaviour originating from the torus amplitude, that after a four-dimensional reduction on a spectator $T^3$ yields the announced $R^{-4}$ term in \vacuum, receives only exponentially suppressed contributions from ${\scr K}$, ${\scr A}$ and ${\scr M}$ provided the radii are appropriately correlated \suppressing.

This is the simplest instance of a class of non-supersymmetric orientifolds with two large transverse dimensions and a naturally small cosmological constant. In the large-radius limit supersymmetry is restored in the bulk, while the D-brane spectra stay non-supersymmetric, but exhibit Fermi-Bose degeneracy at all massive string levels. In loose terms, the model contains two ``mirror worlds'', and the degeneracy is due to an interchange of the ordinary superpartners on the two branes.

\newsec{Taming higher-order corrections}

Having found a model with a Fermi-Bose degenerate massless spectrum in the open-string sector does not suffice, however, to guarantee that higher-loop corrections do not induce sizeable contributions to the vacuum energy. In orientifold models these originate 
by diagrams with increasing numbers of handles, crosscaps and holes, and are in general quite hard to compute directly (see, for instance, \higher). Despite these technical difficulties it is at times possible to get a flavour of the qualitative behaviour of higher-genus vacuum amplitudes and to discriminate between those that are vanishing identically and those that are not. 

Let us consider for example the model presented in \cardella. It consists essentially in a discrete deformation of the open-string sector allowed by the two-dimensional CFT constraints \yassen, in such a way that a fully supersymmetric bulk is accompanied by non-supersymmetric branes with gauge group ${\rm SO} (8) \times {\rm USp} (8)$ and fermions in the $(36,1)+(1,28)$. The relevant transverse-channel amplitudes for our discussion are (see \cardella\ for more details)
\eqn\Klein{
\Kt = {2^4 \over 2} (V_8 - S_8 ) \, (O_4 O_4 + V_4 V_4) \,,
}
\eqn\annulus{
\eqalign{
\At &= {2^{-4}\over 2}\, \left\{ (N+M)^2 \, (V_8 - S_8 ) \,  
\left( O_4 O_4 + V_4 O_4 + S_4 S_4 + C_4 S_4 \right) \right.
\cr
& \left. + \left[ (N-M)^2 \, V_8 - (-N+M)^2\, S_8 \right] \left( V_4 V_4 + O_4 V_4 + C_4 C_4 + S_4 C_4 \right) 
\right\}\,,
\cr}
}
and
\eqn\Moebius{
\Mt = - \left\{ (N+M) \, (\hat V_8 - \hat S_8) \, \hat O_4 \hat O_4 +
\left[ (N-M)\, \hat V_8 - (-N+M)\,  \hat S_8\right] \, \hat V_4 \hat V_4 \right\} \,.
}
It is then clear that for $N=M$ (both equal to eight as a result of tadpole conditions) these one loop amplitudes vanish identically, even in the presence of supersymmetry breaking, as a result of the Jacobi identity $V_8 \equiv S_8$. 

Moving to higher-genus, the amplitudes associated with closed Riemann surfaces, both oriented and unoriented, are expected to vanish, since the closed-string sector is not affected by the deformation and hence have the same properties as the 
type I superstring. However, more care is needed when surfaces with boundaries are considered. Let us specialise to surfaces with two crosscaps and one boundary. Similarly to the one-loop case, there is a particular choice for the period matrix $\varOmega_{\alpha\beta}$ for which this surface describes a tree-level three-closed-string interaction diagram, weighted by the product of disc ($B_i$) and cross-cap ($\varGamma_i$) one-point functions of closed states, that can be read from the transverse-channel Klein-bottle, annulus and M\"obius-strip amplitudes. More precisely, 
the expression for the amplitude would be
$$
{\scr R} _{[0,1,2]} = \sum_{i,j,k} \varGamma_i\, \varGamma_j \,B_k \, {\scr N}_{ij}{}^k \, {\scr V}_{ij}{}^k (\varOmega_{\alpha\beta}) \,,
$$
where the ${\scr N}_{ij}{}^k$ are the fusion rule coefficients and $ {\scr V}_{ij}{}^k (\varOmega_{\alpha\beta})$ are complicated functions of the period matrix encoding the three-point interaction among states $i$, $j$ and $k$. 
This amplitude is expected to vanish in the supersymmetric (undeformed) case, and this requirement imposes  some relations among the functions ${\scr V}\,{}_{ij}{}^k (\varOmega_{\alpha\beta} )$ that we have not defined explicitly. For instance, for the undeformed supersymmetric version of the model in \annulus\ and \Moebius , whose open-string amplitudes are now
$$
\eqalign{
\At &= {2^{-4}\over 2}\, \left[ (N+M)^2 \, (V_8 - S_8 ) \,  
\left( O_4 O_4 + V_4 O_4 + S_4 S_4 + C_4 S_4 \right) \right.
\cr
&  \left. + (N-M)^2 \, (V_8 -  S_8) \left( V_4 V_4 + O_4 V_4 + C_4 C_4 + S_4 C_4 \right) 
\right]
\cr}
$$
and
$$
\Mt = - (N+M) \, (\hat V_8 - \hat S_8) \, \hat O_4 \hat O_4 - (N-M)\, (\hat V_8 -   
\hat S_8) \, \hat V_4 \hat V_4 \,,
$$
the genus three-half amplitude takes the form
\eqn\thsusy{
\eqalign{
{\scr R}_{[0,1,2]} &= {\textstyle{1\over 4}} \, (N+M) \, \left[ {\scr V}\,{}_{111} + 
3 {\scr V}\,{}_{133} + {\scr V}\,{}_{122} + {\scr V}\,{}_{144} +2{\scr V}\,{}_{234} \right]
\cr
&+ {\textstyle{1\over 4}} \, (N-M) \, \left[
{\scr V}\,{}_{122} + {\scr V}\,{}_{144} +2{\scr V}\,{}_{234} \right] \,,
\cr}
}
where the relative numerical coefficients of the ${\scr V}\,\,$'s take into account the combinatorics of diagrams with given external states. The indices $1,2,3,4$ refer to the four characters $V_8\, O_4 O_4$,
$V_8 \, V_4 V_4$, $-S_8 \, O_4 O_4$ and $-S_8 \, V_4 V_4$, that identify the only states  with a non-vanishing $\varGamma_i$, as can be read from eq. \Klein, and
their non-vanishing fusion rule coefficients, all equal to one, are ${\scr N}\,{}_{111}$, ${\scr N}\,{}_{122}$, ${\scr N}\,{}_{133}$, ${\scr N}\,{}_{144}$ and ${\scr N}\,{}_{234}$. Finally, both in ${\scr V}\,$ and in ${\scr N}\,$ we have lowered the indices using the diagonal metric $\delta_{kl}$, since all characters in this model are self conjugate.

For a supersymmetric theory this amplitude is expected to vanish independently of brane locations, and thus the condition ${\scr R}_{[0,1,2]} = 0$ amounts to the two constraints
\eqn\relations{
 \eqalign{
 {\scr V}\,{}_{111} + 3 {\scr V}\,{}_{133} =& 0 \,,
\cr
{\scr V}\,{}_{122} + {\scr V}\,{}_{144} +2{\scr V}\,{}_{234} =& 0 \,.
\cr}
}
Turning to the non-supersymmetric open sector in eqs. \annulus\ and \Moebius, one finds instead
\eqn\thnonsus{
\eqalign{
{\scr R}_{[0,1,2]} &= {\textstyle{1\over 4}} \, (N+M) \, \left[ {\scr V}\,{}_{111} + 
3 {\scr V}\,{}_{133} + {\scr V}\,{}_{122} + {\scr V}\,{}_{144} +2{\scr V}\,{}_{234} \right]
\cr
&- {\textstyle{1\over 2}} \, (N-M) \, \left[
{\scr V}\,{}_{122} - {\scr V}\,{}_{144} \right] \,,
\cr}
}
since now $B_4 = -N +M$ has a reversed sign. Using eq. \relations\ the 
non-vanishing contribution to the genus three-half vacuum energy would be
$$
{\scr R}_{[0,1,2]} = - {\textstyle{1\over 2}} \, (N-M) \, \left[
{\scr V}\,{}_{122} - {\scr V}\,{}_{144} \right]
$$
that however vanishes for our choice $N=M$.

Similar considerations hold for other genus-$g$ surfaces with boundaries. In all cases it can be shown that all the potentially non-vanishing contributions are multiplied by the breaking coefficients $N-M$, and thus are zero for $M=N$. As a result, no contributions to the vacuum energy are generated at any order in perturbation theory. Unfortunately this configuration of orientifold planes and D-branes is unstable, and in the true vacuum already the one loop amplitudes contribute to $\varLambda$.

\vskip 24pt
\noindent
{\bf Acknowledgement} I would like to thank Prof. Eugenio Coccia and the Board of the Italian Society of General Relativity and Gravitation for selecting me for the 2004 SIGRAV prize.
It is amusing to notice a nice coincidence of dates: this 2004 prize came ten years later my supervisors had been awarded the first SIGRAV prize for the ``systematics of open-string constructions'', while I was about to start my PhD under their supervision.
It is a real pleasure to acknowledge Adi Armoni, Massimo Bianchi, Ralph Blumenhagen, Matteo Cardella, Giuseppe D'Appollonio, Riccardo D'Auria, Emilian Dudas, Sergio Ferrara, Kristin F\"orger, Matthias Gaberdiel, Jihad Mourad, Gianfranco Pradisi, Yassen S. Stanev, Mario Trigiante and especially Ignatios Antoniadis and Augusto Sagnotti for their friendship and for long-lasting stimulating collaborations that have contributed significantly to shape my understanding of string theory. Ideally, I would like to share the SIGRAV prize with them all.

\listrefs

\end